# Test of mode-division multiplexing and demultiplexing in free-space with diffractive transformation optics

GIANLUCA RUFFATO,[1,2,*] MICHELE MASSARI,[1,2] GIUSEPPE PARISI,[3] AND FILIPPO ROMANATO[1,2,4]

[1]Department of Physics and Astronomy 'G. Galilei' ,University of Padova, via Marzolo 8,35131 Padova, Italy
[2]Laboratory for Nanofabrication of Nanodevices, c.so Stati Uniti 4, 35127 Padova, Italy
[3]SM Optics – SIAE Group, Via M. Buonarroti 21, 20093 Cologno Monzese, Milano, Italy
[4]CNR -INFM TASC IOM National Laboratory, S.S. 14 Km 163.5, 34012 Basovizza, Trieste, Italy
*Corresponding author: gianluca.ruffato@unipd.it

**In recent years, mode-division multiplexing (MDM) has been proposed as a promising solution in order to increase the information capacity of optical networks both in free-space and in optical fiber transmission. Here we present the design, fabrication and test of diffractive optical elements for mode-division multiplexing based on optical transformations in the visible range. Diffractive optics have been fabricated by means of 3D high-resolution electron beam lithography on polymethylmethacrylate resist layer spun over a glass substrate. The same optical sequence was exploited both for input-mode multiplexing and for mode sorting after free-space propagation. Their high miniaturization level and efficiency make these optical devices ideal for integration into next-generation platforms for mode-division (de)multiplexing in telecom applications.**

*OCIS codes:* (050.1970) Diffractive optics; (050.4865) Optical vortices; (060.4510) Optical communications; (110.6895) Three-dimensional lithography

In the last decades, several methods have been presented in order to deal with the growing worldwide demand of bandwidth and to boost the information capacity of optical networks [1]. Basically, almost all these techniques were about the manipulation of different physical dimensions of light waves, including frequency/wavelength, time, complex amplitude and polarization. More recently, the attention has been focused on the spatial degree of freedom, in the so-called spatial division multiplexing (SDM), consisting in tailoring the spatial structure and distribution of the transmitted waves. Mode-division multiplexing (MDM), in particular, is aimed at exploiting the several orthogonal modes supported by the transmission medium as distinct information channels. The axial symmetry of optical fibers and, obviously, of free-space, suggests the selection of modes carrying orbital angular momentum (OAM) of light as possible candidates [2]. These modes are characterized by a helical phase form $\exp(i\ell\varphi)$ ($\ell=0,\pm1,\pm2,...$), being $\ell$ the topological charge and $\varphi$ the azimuthal coordinate [3, 4]. Beams carrying different OAM are intrinsically orthogonal and separable with each other. The exploitation of OAM modes has demonstrated to allow promising results both in free-space [5, 6] and in optical fiber transmission [7], either as distinct information channels or as high-dimensional alphabet for classical [8] and quantum applications [9].

The crucial parts of an optical link based on MDM are represented by the multiplexer and the demultiplexer, that is by the optical techniques exploited to prepare a collimated superposition of modes at the transmitter and to separate them at the receiver, respectively. Different solutions have been presented and described in order to sort a set of multiplexed beams differing in their OAM content: interferometric methods [10], optical transformations [11-18], time-division technique [19], integrated silicon photonics [20], coherent detection [21], OAM-mode analyzers [22-24].

These methods are commonly presented and considered for demultiplexing operations, and multiplexing is only marginally demonstrated by invoking the invariance of the light path for time-reversal [25]. Therefore, very little attention has been devoted in literature to the multiplexing process, which is usually performed with cumbersome beam-splitters, in a lossy and non-scalable manner. Recently, mode multiplexers based on multiplane light conversion [26], complex phase mask and gratings [27, 28], q-plates [29], fiber and photonic integrated devices [30, 31] have been proposed as promising candidates for optical vortex multiplexing.

Phase-only diffractive optical elements (DOE) appear as the most suitable choice for the realization of passive and lossless, compact and cheap optical devices for integrated (de)multiplexing applications. Among all the sorting techniques, OAM-mode analyzers and transformation optics appear the most suitable to be realized in a diffractive form. While OAM-mode analyzers allow performing demultiplexing with a single optical element and a high freedom in the far-field channel constellation design [32, 33], on the other hand their efficiency is inversely proportional to the number of states being sampled, which makes this technique nonviable especially for quantum applications. Due to their higher efficiency, we considered the design, fabrication and characterization of diffractive elements based on transformation optics and we tested in sequence both the multiplexing and demultiplexing optical processes at $\lambda$=632.8 nm.

Solutions based on this technique demonstrated how OAM states can be efficiently converted into transverse momentum states through a *log-pol* optical transformation [11]. Two elements are necessary: the unwrapper and the phase-corrector. The former performs the conformal mapping of a position $(x, y)$ in the input plane to a position $(u, v)$ in the output plane, where $v=a \arctan(y/x)$ and $u= -a \ln(r/b)$, being $r=(x^2+y^2)^{1/2}$, $a$ and $b$ design parameters. Its phase function $\Omega_1$ is:

$$\Omega_1(x, y) = \frac{2\pi a}{\lambda f_1}\left[ y \arctan\left(\frac{y}{x}\right) - x \ln\left(\frac{\sqrt{x^2+y^2}}{b}\right) + x + \frac{x^2+y^2}{2a}\right] \quad (1)$$

The two free parameters $a$ and $b$ determine the scaling and position of the transformed beam respectively. The parameter $a$ takes the value $L/2\pi$, ensuring that the azimuthal angle range $(0, 2\pi)$ is mapped onto a length $L$ which is shorter than the full width of the second element. The parameter $b$ is optimized for the particular physical dimensions of the sorter and can be chosen independently. The phase-corrector, placed at a distance $f_1$, has a phase function $\Omega_2$ given by:

$$\Omega_2(u, v) = -\frac{2\pi ab}{\lambda f_1}\exp\left(-\frac{u}{a}\right)\cos\left(\frac{v}{a}\right) \quad (2)$$

A lens with focal length $f_2$ is inserted after the phase-corrector element in order to focus the transformed beam onto a specified lateral position, which moves proportionally to the OAM content $\ell$ according to:

$$\Delta s = \frac{f_2 \lambda}{2\pi a}\ell \quad (3)$$

Alternatively, the focusing quadratic term can be integrated in the phase-corrector element as well.

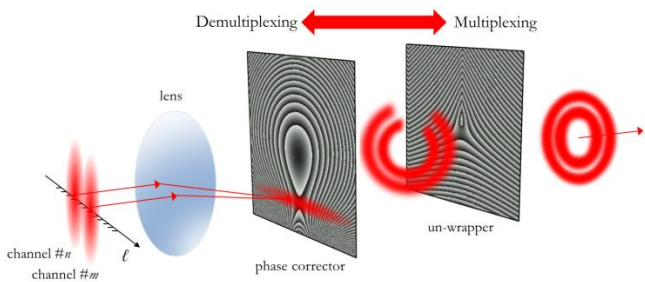

Fig. 1. Schematic of the (de)multiplexer working principle. From left to right: multiplexing. The input beam should be reshaped into an elongated form, for instance using a cylindrical lens. Then it is Fourier-transformed with a lens and it illuminates the first optical element with a tilt angle, depending on the initial axial displacement with respect to the axis of the lens. Then the beam is wrapped and illuminates the second optical element for phase correction. The system converts an input linear phase gradient, created by the tilted incidence, into an output azimuthal phase gradient. Beams with different axial displacements in input are converted into beams with different OAM content. From right to left: demultiplexing.

The same setup has been demonstrated to work as multiplexer, in reverse. In this case the two elements are illuminated in the opposite direction. The input beam should be properly reshaped into an asymmetric elongated spot which is wrapped by the first element (phase corrector) and then corrected in its phase by the second element (unwrapper). Thus the azimuthal phase gradient of the output beam is achieved by wrapping the input linear phase gradient, which is in turn obtained by illuminating the first element with a non-null incidence angle. In [17] for instance, this was obtained by using the diffraction pattern of an axially-shifted slit. Here we propose a more efficient technique, similar to the one exploited in [16], consisting in reshaping a Gaussian beam with a cylindrical lens and using a spherical lens in *f-f* configuration to convert the axial displacement of the input beam into a tilted one, thanks to Fourier transform properties.

As far as fabrication is concerned, in its first realization the two elements were implemented with spatial light modulators (SLMs) [11]. At a later stage [12], they were replaced by two freeform refractive optical components, exhibiting higher efficiency, though not a small size. In this work, we further improve the miniaturization level by realizing them in a diffractive form with 256 phase levels.

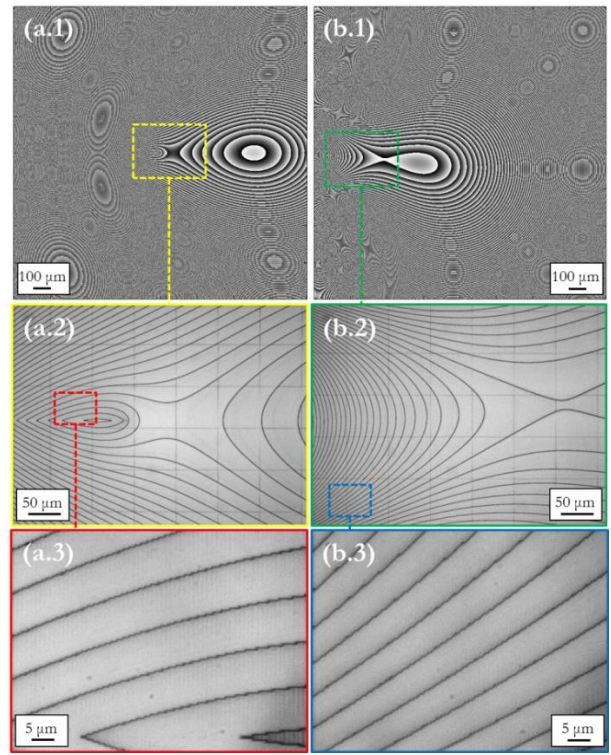

Fig. 2. Diffractive optics for *log-pol* coordinate transformation. Computed phase patterns (a.1, b.1), optical microscope inspections and details: un-wrapper (a.2, a.3) and phase corrector (b.2, b.3). Design parameters: $a$=220 μm, $b$=50 μm, $f_1$=9 mm, carrier spatial frequency in the phase corrector for image tilt (in the demultiplexer): $\alpha=\beta$=0.1 μm$^{-1}$. Optimized for wavelength $\lambda$=632.8 nm. 256 phase levels. Total size: 1.6 x 1.6 mm$^2$.

Diffractive optical elements are fabricated as surface-relief patterns of pixels. This 3-D structure can be realized by shaping a layer of transparent material, imposing a direct proportionality between the thickness of the material and the phase delay. Electron beam lithography (EBL) is the best technique in order to fabricate 3D profiles with high resolution [33, 34]. In this work, the DOE patterns were written on a polymethylmethacrylate (PMMA) resist layer with a JBX-

6300FS JEOL EBL machine, 12 MHz, 5 nm resolution, working at 100 keV with a current of 100pA. The substrate used for the fabrication is glass coated with an ITO layer with conductivity of 8-12 Ω, in order to ensure both transparency and a good discharge during the exposure. Patterned samples were developed under slight agitation in a temperature-controlled developer bath for 60 s in a solution of deionized water: isopropyl alcohol (IPA) 3:7. For the working wavelength $\lambda$=632.8nm, PMMA refractive index results $n_{PMMA}$=1.489 from spectroscopic ellipsometry analysis (J.A. Woollam VASE, 0.3 nm spectral resolution, 0.005° angular resolution). The height $d_k$ of the pixel belonging to the $k$th layer is given by:

$$d_k = \frac{\lambda}{n_{PMMA}-1} \frac{k-1}{N} \quad (4)$$

being $N$ the number of phase levels. In our case of interest, for $N$=256 we get: $d_1$=0 nm, $d_{256}$=1289.4 nm, step $\Delta d$=5.1 nm. In figure 2 fabricated samples are shown, with design parameters $f_1$=9 mm, $a$=220 μm, $b$=50 μm. A tilt term was added to the phase corrector in the demultiplexing sequence, with carrier spatial frequencies $\alpha=\beta$= 0.1 μm$^{-1}$, in order to prevent any overlap on the CCD with the potentially noise-carrier zero-order term.

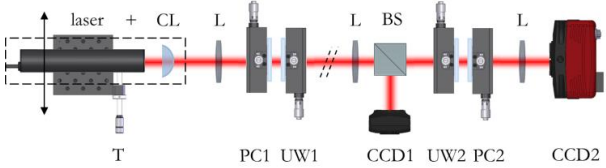

Fig. 3. Experimental setup for OAM sorting with a single diffractive optical element (DOE) implementing transformation optics. The laser illuminates a cylindrical lens for input beam reshaping. The input system can move perpendicularly to the propagation direction with a micrometric translator (T). Then the beam is Fourier transformed with a lens and illuminates the sequence of optical elements PC1-UW1 performing multiplexing. After travelling in free-space, the generated vortex is split with a 50:50 beam-splitter (BS) for beam analysis with the first camera (CCD1). The second part of the beam enters the demultiplexing sequence UW2-PC2. Finally, the beam is Fourier-transformed by a lens (L) and collected on the second camera (CCD2).

The characterization setup was mounted on an optical table (Figure 3). The Gaussian beam ($\lambda$= 632.8 nm, beam waist $w_0$=240 μm, power 0.8 mW) emitted by a HeNe laser source (HNR008R, Thorlabs) is reshaped with a cylindrical lens in order to prepare an elongated input beam to be wrapped by the multiplexer. Both laser and reshaping lens are mounted on the same stage and can be translated together using a micrometric translator (TADC-651, Optosigma) in the direction orthogonal to the propagation direction. The transmitted beam is Fourier-transformed with a first lens of focal length $f_0$=75 mm, which is exploited to convert the axial displacement of the input beam into a tilted beam illuminating the first diffractive element (phase corrector). For increasing laser shift, i.e. increasing angle of incidence on the multiplexer, beams with increasing OAM are generated. By translating the laser in opposite directions with respect to the $\ell$=0 position, beams with opposite helicity are generated (figure 4). Both phase corrector and un-wrapper are mounted on a $XY$ translation sample holder with micrometric drives for sample alignment. A beam-splitter is used to analyze the field profile, which is collected with a CCD camera (DCC1545M, Thorlabs, 1280x1024 pixels, 5.2 μm pixel size, monochrome, 8-bit depth). Then the beam illuminates the demultiplexing sequence. At last, the far-field is collected by a second CCD camera (1500M-GE, Thorlabs, 1392x1040 pixels, 6.45 μm pixel size, monochrome, 12-bit depth) placed at the back-focal plane of a lens of focal length $f_2$=10 cm. In far-field, a bright asymmetric spot appears whose position shifts linearly with the OAM content, i.e. with the laser lateral position (figure 4.d). In figure 4.b the output field is shown for an input beam corresponding to $\ell$=-4. Subsequently, a further laser beam was added, placing a second beam-splitter before the multiplexing sequence. The second laser is mounted on a linear translator as well, and its OAM content can be controlled independently from the first one. In figure 4.c the two multiplexed beams are shown in case of $\ell$ =−5 and $\ell$=+5 respectively. As expected, while the beams are superimposed after the multiplexer, two clearly separated signals appear after demultiplexing.

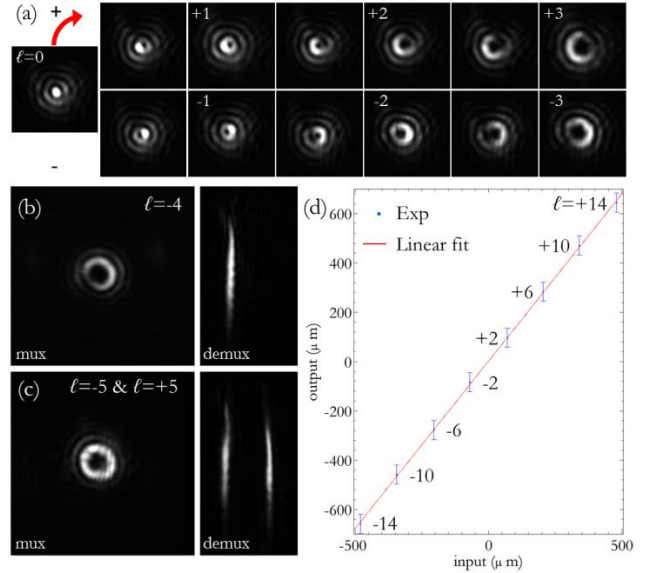

Fig. 4. (a) Images recorded with a CCD camera are shown for various OAM states up to the 3rd order (white numbers), for translation of the laser towards opposite directions. The distance between the lateral positions of the laser that generate integer OAM states was around 35 μm. In between these positions, fractional OAM modes are generated. (b) Generation and detection of a beam corresponding to $\ell$=-4. (c) Multiplexing and demultiplexing of two beams corresponding to $\ell$=+5 and $\ell$=-5. (d) Output position as a function of input laser shift, experimental data for $\ell$ in the range {-14,…,+14}, step $\Delta\ell$=4

The CCD illuminated area is split into rectangular regions which are centered on each elongated spot in far-field with a size corresponding to the minimum separation between any two adjacent channels. We defined 29 of these regions, and we analyzed the optical response under illumination with pure vortices with $\ell$ values spanning in the range from $\ell$=−14 to $\ell$=+14. As highlighted elsewhere [11-13], a limitation of OAM-beam sorting with *log-pol* transformation is represented by the slightly overlap between adjacent channels, which results into detrimental inter-channel cross-talk. One option for improvement is to include a fan-out element [14, 15], which extends the phase gradient by introducing multiple copies, thus providing a larger separation between spots, at the expense of increased complexity and size of the optical system. Alternatively, the choice of non-consecutive OAM values can further diminish channel cross-talk. The cross-talk $XT$ on the channel $\ell=\ell^*$ is defined as:

$$XT_{\ell=\ell^*} = 10 \cdot \log_{10} \frac{I_{\ell^*,ALL\setminus\ell^*}}{I_{\ell^*,ALL}} \quad (5)$$

where $I_{\ell^*\text{ALL}}$ is the signal in correspondence of channel $\ell^*$ when all input OAM signals in the set $\{\ell_i\}$ are on, $\ell^*$ included, while $I_{\text{ALL}\setminus\{\ell^*\}}$ is the signal at channel $\ell^*$ when the input channel $\ell^*$ is off. The choice of $\Delta\ell=4$ provides a good separation of far-field spots with channel efficiencies up to 94% and values lower than 4% for off-diagonal terms (figures 5 and 6.a) and allows obtaining acceptable cross-talk values below -15 dB (figure 6.b). The separation between consecutive spots is around 180.6 μm, close to the theoretical value 183.2 μm calculated with eq. (3).

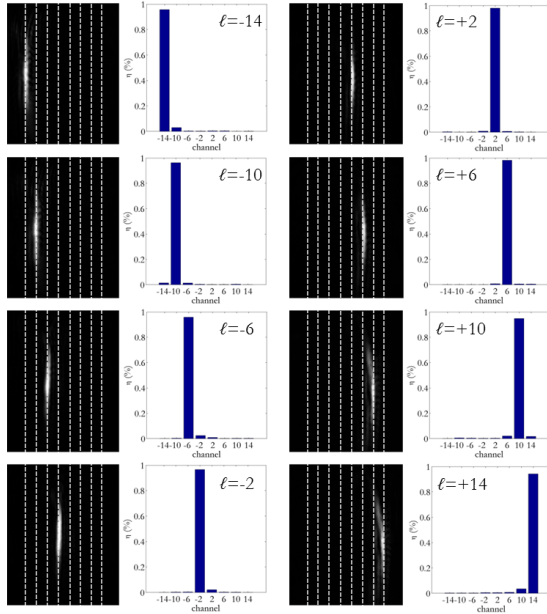

Fig. 5. Experimental intensity output and efficiency for 8 input modes, in the range $\ell=\{-14,…,+14\}$, step $\Delta\ell=4$. Bar-plot: channels efficiency.

In conclusion, we fabricated phase-only diffractive optical elements with high-resolution electron beam lithography and we demonstrated their performance for mode-division (de)multiplexing based on transformation optics. The fabricated elements were integrated into a prototypal free-space link and the multiplexing and demultiplexing optical processes were tested using the same optical sequence, in reverse. By properly choosing the design parameters and the channels set, cross-talk values below -15 dB can be achieved. Thanks to the high efficiency and miniaturization level, the fabricated optics are promising for integration into optical platforms performing optical processing of OAM channels, also for applications in optical fibers.

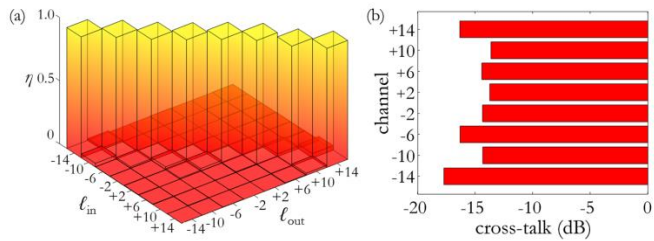

Fig. 6. Efficiency (a) and cross-talk (b) for 8 channels in the set $\{-14, …, +14\}$, step $\Delta\ell=4$.

## REFERENCES

1. E. Agrell, M. Karlsson, A. R. Chraplyvy, D. J. Rochardson, P. M. Krummrich, P. Winzer, K. Roberts, J. K. Fisher, S. J. Savory, B. J. Eggleton, M. Secondini, F. R. Kschischang, A. Lord, J. prat, I. Tomkos, J. E. Bowers, S. Srinivasan, M. Brandt-Pearce, and N. Gisin, *J. Optics* **18**, 063002 (2016).
2. S. Yu, *Opt. Express* **23**, 3075 (2015).
3. L. Allen, M. W. Beijersbergen, R. J. C. Spreeuw, and J. P. Woerdman, *Phys. Rev. A* **45**, 8185 (1992).
4. D. L. Andrews and M. Babiker, The Angular Momentum of Light (Cambridge University Press, 2013).
5. J. Wang, J.-Y. Yang, I. M. Fazal, N. Ahmed, Y. Yan, H. Huang, Y. Ren, Y. Yue, S. Dolinar, M. Tur, and A. E. Willner, *Nat. Phot.* **6**, 488 (2012).
6. F. Tamburini, E. Mari, A. Sponselli, B. Thidé, A. Bianchini, and F. Romanato, New. J. Phys. 14, 033001 (2012).
7. N. Bozinovic, Y. Yue, Y. Ren, N. Tur, P. Kristensen, H. Huang, A. E. Willner, and S. Ramachandran, *Science* **340**, 1545 (2013).
8. A. Trichil, C. Rosales-Guzman, A. Dudley, B. Ndagano, A. B. Salem, M. Zghal and A. Forbes, *Sci. Rep.* **6**, 27674 (2016).
9. M. Mirhosseini, O. S. Magana-Loaiza, M. N. O'Sullivan, B. Rudenburg, M. Malik, M. P. J. Lavery, M. J. Padgett, D. J. Gauthier, and R. W. Boyd, *New J. Phys.* **17**, 033033 (2015).
10. J. Leach, M. J. Padgett, S. M. Barnett, S. Franke-Arnold, and J. Courtial, *Phys. Rev. Lett.* **88**, 257901 (2002).
11. G. C. G. Berkhout, M. P. J. Lavery, J. Courtial, M. W. Beijersbergen, and M. J. Padgett, *Phys. Rev. Lett.* **105**, 153601 (2010).
12. M. P. J. Lavery, D. J. Robertson, G. C. G. Berkhout, G. D. Love, M. J. Padgett, and J. Courtial, *Opt. Express* **20**, 2110 (2012).
13. M. P. J. Lavery, D. J. Robertson, A. Sponselli, J. Courtial N. K. Steinhoff, G. A. Tyler, A. E. Willner and M. J. Padgett, *New J. Phys.* **15**, 013024 (2013).
14. M. N. O. Sullivan, M. Mirhosseini, M. Malik, and R. W. Boyd, *Opt. Express* **20**, 24444 (2012).
15. M. Mirhosseini, M. Malik, Z. Shi and R. W. Boyd, *Nat. Comm.* **4**, 2781 (2013).
16. H. Huang, G. Milione, M. P. J. Lavery, G. Xie, Y. Ren, Y. Cao, N. Ahmed, T. A. Nguyen, D. A. Nolan, M.-J. Li, M. Tur, R. R. Alfano and A. E. Willner, *Sci. Rep.* **5**, 14931 (2015).
17. R. Fickler, R. Lapkiewicz, M. Huber, M. P. J. Lavery, M. J. Padgett and A. Zeilinger, *Nat. Comm.* **5**, 4502 (2014).
18. K. S. Morgan, I. S. Raghu, E. G. Johnson, *Proc. of SPIE* **9374**, 93740Y (2015).
19. P. Bierdz, M. Kwon, C. Roncaioli, and H. Deng, *New J. Phys.* **15**, 113062 (2013).
20. T. Su, R. P. Scott, S. S. Djordjevic, N. K. Fontaine, D. J. Geisler, X. Cai, and S. J. B. Yoo, *Opt. Express* **20**, 9396 (2012).
21. A. Belmonte, and J. P. Torres, *Opt. Lett.* **38**, 241 (2013).
22. V. V. Kotlyar, S. N. Khonina, and V. A. Soifer, *J. Mod. Opt.* **45**, 1495 (1998).
23. G. Gibson, J. Courtial, M. J. Padgett, M. Vasnetsov, V. Pas'ko, S. M. Barnett, and S. Franke-Arnold, *Opt. Express* **12**, 5448 (2004).
24. N. Zhang, X. C. Yuan, and R. E. Burge, *Opt. Lett.* **35**, 3495 (2010).
25. J. Wang, *Photon. Res.* **4**, B14 (2016).
26. G. Labroille, B. Denolle, P. Jian, P. Genevaux, N. Treps, and J.-F. Morizur, *Opt. Express* **22**, 15599 (2014).
27. S. Gao, T. Lei, Y. Li, Y. Yuan, Z. Xie, Z. Li, and X. Yuan, *Opt. Express* **24**, 21642 (2016).
28. T. Lei, M. Zhang, Y. Li, P. Jia, G. N. Liu, X. Xu, Z. Li, C. Min, J. Lin, C. Yu, H. Niu, and X. Yuan, *Light: Science and Applications* **4**, e257 (2015).
29. G. Milione, M. P. J. Lavery, H. Huang, Y. Ren, G. Xie, T. A. Nguyen, E. Karimi, L. Marrucci, D. A. Nolan, R. R. Alfano, and A. E. Willner, *Opt. Lett.* **40**, 1980 (2015).
30. Y. Yan, Y. Yue, H. Huang, J. Y. Yang, M. R. Chitgarha, N. Ahmed, M. Tur, S. J. Dolinar, and A. E. Willner, *Opt. Lett.* **37**, 3645 (2012).
31. B. Guan, R. P. Scott, C. Qin, N. K. Fontaine, T. Su, C. Ferrari, M. Cappuzzo, F. Klemens, B. Keller, M. Earnshaw, and S. J. B. Yoo, *Opt. Express* **22**, 145 (2014).
32. S. Li and J. Wang, *Sci. Rep*. **5**, 15406 (2015).
33. G. Ruffato, M. Massari and F. Romanato, *Sci. Rep.* **6**, 24760 (2016).
34. M. Massari, G. Ruffato, M. Gintoli, F. Ricci, and F. Romanato, *App. Opt.* **54**, 4077 (2015).

# REFERENCES

1. E. Agrell, M. Karlsson, A. R. Chraplyvy, D. J. Rochardson, P. M. Krummrich, P. Winzer, K. Roberts, J. K. Fisher, S. J. Savory, B. J. Eggleton, M. Secondini, F. R. Kschischang, A. Lord, J. prat, I. Tomkos, J. E. Bowers, S. Srinivasan, M. Brandt-Pearce, and N. Gisin, "Roadmap of optical communications", *J. Optics* **18**, 063002 (2016).
2. S. Yu, "Potential and challenges of using orbital angular momentum communications in optical interconnects", *Opt. Express* **23**, 3075-3087 (2015).
3. L. Allen, M. W. Beijersbergen, R. J. C. Spreeuw, and J. P. Woerdman, "Orbital angular momentum of light and the transformation of Laguerre-Gaussian modes", *Phys. Rev. A* **45**, 8185-8189 (1992).
4. D. L. Andrews and M. Babiker, The Angular Momentum of Light (Cambridge University Press, 2013).
5. J. Wang, J.-Y. Yang, I. M. Fazal, N. Ahmed, Y. Yan, H. Huang, Y. Ren, Y. Yue, S. Dolinar, M. Tur, and A. E. Willner, "Terabit free-space data transmission employing orbital angular momentum multiplexing", *Nat. Phot.* **6**, 488-496 (2012).
6. F. Tamburini, E. Mari, A. Sponselli, B. Thidé, A. Bianchini, and F. Romanato, "Encoding many channels on the same frequency through radio vorticity: first experimental test", New. J. Phys. 14, 033001-1-17 (2012).
7. N. Bozinovic, Y. Yue, Y. Ren, N. Tur, P. Kristensen, H. Huang, A. E. Willner, and S. Ramachandran, "Terabit-scale orbital angular momentum mode division multiplexing in fibers", *Science* **340**, 1545-1548 (2013).
8. A. Trichil, C. Rosales-Guzman, A. Dudley, B. Ndagano, A. B. Salem, M. Zghal and A. Forbes, "Optical communication beyond orbital angular momentum", *Sci. Rep.* **6**, 27674 (2016).
9. M. Mirhosseini, O. S. Magana-Loaiza, M. N. O'Sullivan, B. Rudenburg, M. Malik, M. P. J. Lavery, M. J. Padgett, D. J. Gauthier, and R. W. Boyd, "High-dimensional quantum cryptography with twisted light", *New J. Phys.* **17**, 033033-1-12 (2015).
10. J. Leach, M. J. Padgett, S. M. Barnett, S. Franke-Arnold, and J. Courtial, "Measuring the orbital angular momentum of a single photon", *Phys. Rev. Lett.* **88**, 257901-1-4 (2002).
11. G. C. G. Berkhout, M. P. J. Lavery, J. Courtial, M. W. Beijersbergen, and M. J. Padgett, "Efficient sorting of orbital angular momentum states of light", *Phys. Rev. Lett.* **105**, 153601-1-4 (2010).
12. M. P. J. Lavery, D. J. Robertson, G. C. G. Berkhout, G. D. Love, M. J. Padgett, and J. Courtial, "Refractive elements for the measurements of the orbital angular momentum of a single photon", *Opt. Express* **20**, 2110-2115 (2012).
13. M. P. J. Lavery, D. J. Robertson, A. Sponselli, J. Courtial N. K. Steinhoff, G. A. Tyler, A. E. Willner and M. J. Padgett, "Efficient measurement of an optical orbital-angular-momentum spectrum comprising more than 50 states", *New J. Phys.* **15**, 013024 (2013).
14. M. N. O. Sullivan, M. Mirhosseini, M. Malik, and R. W. Boyd, "Near-perfect sorting of orbital angular momentum and angular position states of light", *Opt. Express* **20**, 24444-24449 (2012).
15. M. Mirhosseini, M. Malik, Z. Shi and R. W. Boyd, "Efficient separation of the orbital angular momentum eigenstates of light", *Nat. Comm.* **4**, 2781 (2013).
16. H. Huang, G. Milione, M. P. J. Lavery, G. Xie, Y. Ren, Y. Cao, N. Ahmed, T. A. Nguyen, D. A. Nolan, M.-J. Li, M. Tur, R. R. Alfano and A. E. Willner, "Mode division multiplexing using an orbital angular momentum sorter and MIMO-DSP over a graded-index few-mode optical fibre", *Sci. Rep.* **5**, 14931 (2015).
17. R. Fickler, R. Lapkiewicz, M. Huber, M. P. J. Lavery, M. J. Padgett and A. Zeilinger, "Interface between path and orbital angular momentum entanglement for high-dimensional photonic quantum information", *Nat. Comm.* **5**, 4502 (2014).
18. K. S. Morgan, I. S. Raghu, E. G. Johnson, "Design and fabrication of diffractive optics for orbital angular momentum space division multiplexing", *Proc. of SPIE* **9374**, 93740Y-1-6 (2015).
19. P. Bierdz, M. Kwon, C. Roncaioli, and H. Deng, "High fidelity detection of the orbital angular momentum of light by time mapping", *New J. Phys.* **15**, 113062-1-13 (2013).
20. T. Su, R. P. Scott, S. S. Djordjevic, N. K. Fontaine, D. J. Geisler, X. Cai, and S. J. B. Yoo, "Demonstration of free space coherent optical communication using integrated silicon photonic orbital angular momentum devices", *Opt. Express* **20**, 9396-9402 (2012).
21. A. Belmonte, and J. P. Torres, "Digital coherent receiver for orbital angular momentum demultiplexing", *Opt. Lett.* **38**, 241-243 (2013).
22. V. V. Kotlyar, S. N. Khonina, and V. A. Soifer, "Light field decomposition in angular harmonics by means of diffractive optics", *J. Mod. Opt.* **45**, 1495-1506 (1998).
23. G. Gibson, J. Courtial, M. J. Padgett, M. Vasnetsov, V. Pas'ko, S. M. Barnett, and S. Franke-Arnold, "Free-space information transfer using light beams carrying orbital angular momentum", *Opt. Express* **12**, 5448-5456 (2004).
24. N. Zhang, X. C. Yuan, and R. E. Burge, "Extending the detection range of optical vortices by Dammann vortex gratings", *Opt. Lett.* **35**, 3495-3497 (2010).
25. J. Wang, "Advances in communications using optical vortices", *Photon. Res.* **4**, B14-B28 (2016).
26. G. Labroille, B. Denolle, P. Jian, P. Genevaux, N. Treps, and J.-F. Morizur, "Efficient and mode selective spatial mode multiplexer based on multi-plane light conversion", *Opt. Express* **22**, 15599-15607 (2014).
27. S. Gao, T. Lei, Y. Li, Y. Yuan, Z. Xie, Z. Li, and X. Yuan, "OAM-labeled free-spaced optical flow routing", *Opt. Express* **24**, 21642-21651 (2016).
28. T. Lei, M. Zhang, Y. Li, P. Jia, G. N. Liu, X. Xu, Z. Li, C. Min, J. Lin, C. Yu, H. Niu, and X. Yuan, "Massive individual orbital angular momentum channels for multiplexing enabled by Dammann gratings", *Light: Science and Applications* **4**, e257 (2015).
29. G. Milione, M. P. J. Lavery, H. Huang, Y. Ren, G. Xie, T. A. Nguyen, E. Karimi, L. Marrucci, D. A. Nolan, R. R. Alfano, and A. E. Willner, "4 x 20 Gbit/s mode division multiplexing over free space using vector modes and a q-plate mode (de)multiplexer", *Opt. Lett.* **40**, 1980-1983 (2015).
30. Y. Yan, Y. Yue, H. Huang, J. Y. Yang, M. R. Chitgarha, N. Ahmed, M. Tur, S. J. Dolinar, and A. E. Willner, "Efficient generation and multiplexing of optical orbital angular momentum modes in a ring fiber by using multiple coherent inputs," *Opt. Lett.* **37**, 3645–3647 (2012).
31. B. Guan, R. P. Scott, C. Qin, N. K. Fontaine, T. Su, C. Ferrari, M. Cappuzzo, F. Klemens, B. Keller, M. Earnshaw, and S. J. B. Yoo, "Free-space coherent optical communication with orbital angular, momentum multiplexing/demultiplexing using a hybrid 3D photonic integrated circuit," *Opt. Express* **22**, 145–156 (2014).
32. S. Li and J. Wang, "Simultaneous demultiplexing and steering of multiple orbital angular momentum modes", *Sci. Rep.* **5**, 15406 (2015).
33. G. Ruffato, M. Massari and F. Romanato, "Diffractive optics for combined spatial- and mode- division multiplexing of optical vortices: design, fabrication and optical characterization", *Sci. Rep.* **6**, 24760 (2016).
34. M. Massari, G. Ruffato, M. Gintoli, F. Ricci, and F. Romanato, "Fabrication and characterization of high-quality spiral phase plates for optical applications", *App. Opt.* **54**, 4077-4083 (2015).